\documentclass[showpacs,preprintnumbers,amsmath,amssymb,twocolumn]{revtex4}
\usepackage{graphicx}
\usepackage{dcolumn}
\usepackage{bm}
\usepackage{amssymb}
\usepackage{array}
\usepackage{longtable}
\usepackage{CJK}

\begin{document}


\title{Random walks in weighted networks with a perfect trap: \\An application of Laplacian spectra}

\author{Yuan Lin}

\author{Zhongzhi Zhang}
\email{zhangzz@fudan.edu.cn}
\homepage{http://www.researcherid.com/rid/G-5522-2011}

\affiliation {School of Computer Science, Fudan University,
Shanghai 200433, China}

\affiliation {Shanghai Key Lab of Intelligent Information
Processing, Fudan University, Shanghai 200433, China}

\begin{abstract}
Trapping processes constitute a primary problem of random walks, which characterize various other dynamical processes taking place on networks. Most previous works focused on the case of binary networks, while there is much less related research about weighted networks. In this paper, we propose a general framework for the trapping problem on a weighted network with a perfect trap fixed at an arbitrary node. By utilizing the spectral graph theory, we provide an exact formula for mean first-passage time (MFPT) from one node to another, based on which we deduce an explicit expression for average trapping time (ATT) in terms of the eigenvalues and eigenvectors of the Laplacian matrix associated with the weighted graph, where ATT is the average of MFPTs to the trap over all source nodes. We then further derive a sharp lower bound for the ATT in terms of only the local information of the trap node, which can be obtained in some graphs. Moreover, we deduce the ATT when the trap is distributed uniformly in the whole network. Our results show that network weights play a significant role in the trapping process. To apply our framework, we use the obtained formulas to study random walks on two specific networks: trapping in weighted uncorrelated networks with a deep trap, the weights of which are characterized by a parameter, and L\'evy random walks in a connected binary network with a trap distributed uniformly, which can be looked on as random walks on a weighted network. For weighted uncorrelated networks we show that the ATT to any target node depends on the weight parameter, that is, the ATT to any node can change drastically by modifying the parameter, a phenomenon that is in contrast to that for trapping in binary networks. For L\'evy random walks in any connected network, by using their equivalence to random walks on a weighted complete network, we obtain the optimal exponent characterizing L\'evy random walks, which have the minimal average of ATTs taken over all target nodes.
\end{abstract}



\date{\today}

\maketitle


\section{introduction}

Trapping is a paradigmatic dynamical process, defined as a kind of random walk with a perfect trap fixed at a given position, absorbing all particles (walkers) that visit it~\cite{Mo69}. As an integral major theme of random walks, the trapping process is relevant to a variety of contexts, including target search~\cite{JaBl01,Sh06}, lighting harvesting in antenna systems~\cite{BaKlKo97,BaKl98,BeHoKo03,BeKo06,Ag11}, energy or exciton transport in polymer systems~\cite{SoMaBl97,BlZu81}, and characterizing similarities between the elements of databases~\cite{FoPiReSa07}. In recent years, trapping problem has drawn continuously increasing attention~\cite{LiJuZh12,HwLeKa12,HwLeKa12E,HwLeKa13}.

The highly desirable quantity of the trapping problem is trapping time, also referred to as mean first-passage time (MFPT)~\cite{Re01,Lo96,MeKl04,BuCa05}. The MFPT for a node $i$ to the trap is the expected time taken by a particle starting from $i$ to first arrive at the trap. The average trapping time (ATT) is defined as the average of MFPTs over all source nodes in the system other than the trap, which is very helpful and important, since it is often used as a quantitative indicator measuring the efficiency of trapping process. Hitherto, extensive works have been devoted to evaluating ATT on various complex networks, such as regular lattices with different dimensions~\cite{Mo69,BaKl98JPC,GLKo05,GLLiYoEvKo06,CaAb08}, treelike fractals~\cite{KaRe89,Ag08,HaRo08,ZhLiZhWuGu09,LiWuZh10,ZhWuCh11}, Sierpinski fractals~\cite{KoBa02PRE,KoBa02IJBC,BeTuKo10}, Cayley trees and Vicsek fractals as models of polymer networks~\cite{WuLiZhCh12,LiZh13}, small-world uniform recursive trees~\cite{ZhQiZhGaGu10,ZhLiLiCh11,LiZh13}, as well as many scale-free graphs~\cite{KiCaHaAr08,ZhQiZhXiGu09,ZhGuXiQiZh09,ZhXiZhLiGu09,AgBu09,ZhLiGaZhGuLi09,TeBeVo09,AgBuMa10,ZhYaGa11,ZhYaLi12,MeAgBeVo12}.
These studies have uncovered different impacts of various structural properties on the behavior of trapping process.

Existing works mainly concentrated on the trapping process on binary (unweighted) networks. However, extensive empirical studies~\cite{BoLaMoChHw06} have shown that real-life networks are characterized by not only topologies but also weights, with the latter reflecting the intensive interactions among different nodes~\cite{BaBaPaVe04,MaAlBa05}. The heterogeneity in the distribution of weights can significantly affect diverse dynamical processes on networks,
such as epidemic spreading~\cite{ChZhGuZh11,YaZh12,ZhChXuFu13}, zero-range processes~\cite{TaLiZh06}, synchronization~\cite{ZhMoKu06,Ko07}, cascading failures~\cite{WaCh08,MiBaJaSa11}, games~\cite{YaWaWuLaWa09,BuTo12}, diffusive processes~\cite{BaPa10}, bimolecular chemical reactions~\cite{KwChKi10}, transportation~\cite{RaGo07}, diffusion annihilation processes~\cite{ZhZhGuZh11}, voter behavior~\cite{BaCaPa11}, traffic fluctuation~\cite{ZhZhZhGuZhCh12}, and so on. Thus far, the impact of weights on trapping in weighted networks has not been well understood, although they are suggested to play a significant role in the trapping process. For example, for trapping in dendrimers, weights can be interpreted as energetic funnel controlling the trapping efficiency~\cite{BaKlKo97}. It is, thus, of utmost importance to perform a comprehensive study of trapping in weighted networks, unveiling the influences of weights on the behavior of trapping.


In this paper, we present a theoretical framework for random walks in a weighted network and perform an in-depth study of trapping in a general weighted network with a deep trap located at any node. First, we derive an analytical expression for MFPT from one node to another in any weighted connected network in terms of the eigenvalues and eigenvectors of a Laplacian matrix for the weighted network. Based on this obtained explicit formula for MFPT, we further derive exact solutions to ATT for two cases of trapping problems, which are also expressed in terms of Laplacian spectra and their associated eigenvectors. In the first case, the deep trap is positioned at any given node in the network; while in the other case, the single trap is distributed uniformly over the whole network. Moreover, for the first case, we provide a tight lower bound for the ATT and show that this bound can be achieved in some graphs.

Our obtained results for random walks depend on network weights, which are general and applicable to all weighted networks. To illustrate the universality, we employ our formulas to study two trapping issues in two different types of weighted networks, that is, trapping in weighted uncorrelated networks with a trap fixed at an arbitrary node and trapping in weighted networks with a trap distributed uniformly. The latter is translated from the L\'evy random walks in binary networks, which are dominated by a tunable exponent. The weights of the weighted uncorrelated networks focused on are controlled by a parameter, by changing which one can guide the trapping behavior. These results are also verified by another analytical approach. In addition, for the whole range of the weight parameter, the dominating scalings for ATT are equivalent to that of the aforementioned lower bound for ATT. For L\'evy random walks with a trap distributed uniformly, we obtain the optimal exponent for which the trapping process is most efficient, since in this case the average of MFPTs over all node pairs is the lowest.


\section{Formulation of trapping in generic weighted networks}

In this section, we present a systematic study of trapping problem in a general weighted network. We will derive an analytical expression for MFPT from one node to another, based on which we obtain a closed-form formula of ATT for trapping with a trap at an arbitrary target and a tight lower bound for the ATT, as well as an expression of the ATT for another trapping problem, in which the trap is distritbuted uniformly among the whole network.

\subsection{Introduction of random walks in weighted networks}

To begin with, we give a brief introduction to random walks in a weighted network $G$ with $N$ nodes~\cite{WuXuWuWa07,FrFR09,ZhShCh13}. Mathematically, topological and weighted properties of a weighted network are reflected in a generalized adjacency matrix (weight matrix) $\bf W$ with the element $w_{ij}$ specifying the weight of the edge connecting nodes $i$ and $j$. We focus on undirected networks having symmetric nonnegative weights, namely $w_{ij}=w_{ji}\geq0$. Particularly, $w_{ij}=w_{ji}=0$ indicates that nodes $i$ and $j$ are not adjacent. In a weighted network $G$, the strength of a node $i$ is defined by $s_i=\sum_{j=1}^N w_{ij}$~\cite{BaBaVe04}, which is $i$th nonzero entry of the diagonal strength matrix ${\bf S} = {\rm diag}(s_1, s_2, \ldots, s_N)$, and the Laplacian matrix of $G$ is defined to be ${\bf L}={\bf S}-{\bf W}$.


Differing from most existing works about random walks on binary networks~\cite{ZhAlHoZhCh11}, random walks occurring on weighted networks are biased. In the process of random walks on a weighted network~\cite{WuXuWuWa07,FrFR09,ZhShCh13}, at each time step, the walker starting from current node $i$ jumps to one of its neighboring nodes $j$ with the probability $p_{ij}=w_{ij}/s_i$, which constitutes the $ij$th entry of transition matrix ${\bf P}={\bf S}^{-1}{\bf W}$ governing the diffusion process. It is evident that the transition matrix ${\bf P}$ is a stochastic matrix. The stationary distribution is~\cite{WuXuWuWa07,FrFR09,ZhShCh13}
\begin{equation}\label{EE01}
\pi=(\pi_1, \pi_2, \cdots, \pi_N)^{\top}=\left(\frac{s_1}{s}, \frac{s_2}{s}, \cdots, \frac{s_N}{s}\right)^{\top},
\end{equation}
where $s$ is the sum of strengths over all nodes, namely $s=\sum_{i=1}^N s_i=\sum_{i=1}^{N}\sum_{j=1}^{N} w_{ij}$. Thus, after sufficiently long time, the probability of the walker being at node $i$ is $\pi_i$, which depends on the ratio of the strength $s_i$ of $i$ to the total strength $s$.

After introducing random walks on a weighted network $G$, in what follows, by using the spectral graph theory, we will provide a theoretical analytical formula for MFPT from one node to anther in the weighted network, which forms the basis for other important results of this work, including ATT to an arbitrary trap and its lower bound, as well as the ATT for trapping when the trap is distributed uniformly.


\subsection{Explicit expression for average trapping time to an arbitrary trap}

Let us consider the trapping problem in the weighted network $G$ with a singe trap located at an arbitrary node, and let $T_{ij}$ denote the MFPT from node $i$ to node $j$. For convenience, we label all the $N$ nodes as $1, 2, \cdots, N$, and without loss of generality, we assume that the trap is placed at node $N$. Then, the ATT to node $N$, denoted as $T_N$, is
\begin{equation}\label{EEA01}
T_N=\frac{1}{N-1}\sum_{i=1}^{N-1} T_{iN}.
\end{equation}

In order to deduce $T_{N}$, we should first derive  an expression for $T_{iN}$. According to the definition of MFPT for random walks, we have $T_{NN}=0$ and for $i\neq N$,
\begin{equation}\label{EE02}
T_{iN}=\sum_{k=1}^{N-1}p_{ik}T_{kN}+1,
\end{equation}
which can be rewritten in matrix form as
\begin{equation}\label{EE03}
{\bf \bar{T}}={\bf \bar{e}} + {\bf \bar{P}}{\bf \bar{T}}={\bf \bar{e}}+{\bf \bar{S}}^{-1}{\bf \bar{W}}{\bf \bar{T}},
\end{equation}
where ${\bf \bar{T}}$ is an ($N-1$)-dimensional vector, whose $i$th element is $T_{iN}$; ${\bf \bar{e}}$ is the ($N-1$)-dimensional unit vector $(1, 1, \cdots, 1)^{\top}$; ${\bf \bar{P}}$, ${\bf \bar{S}}$, and ${\bf \bar{W}}$ are, respectively, the submatrices of ${\bf P}$, ${\bf S}$, and ${\bf W}$ by suppressing the $N$th row and $N$th column. Premultiplying Eq.~(\ref{EE03}) by ${\bf \bar{S}}$, we obtain ${\bf \bar{S}}{\bf \bar{T}}={\bf \bar{W}}{\bf \bar{T}}+{\bf \bar{S}}{\bf \bar{e}}$. Therefore,
\begin{equation}\label{EE04}
({\bf \bar{S}}-{\bf \bar{W}}){\bf \bar{T}}={\bf \bar{S}}{\bf \bar{e}}.
\end{equation}
By defining ${\bf \bar{L}} = {\bf \bar{S}}-{\bf \bar{W}}$, which is actually a submatrix of ${\bf L}$ by deleting the $N$th row and $N$th column, Eq.~(\ref{EE04}) can be recast as
\begin{equation}\label{EE05}
{\bf \bar{T}}={\bf \bar{L}}^{-1}{\bf \bar{S}}{\bf \bar{e}},
\end{equation}
implying that
\begin{equation}\label{EE06}
T_{iN}=\sum_{k=1}^{N-1}\bar{l}^{-1}_{ik}s_k,
\end{equation}
where $\bar{l}^{-1}_{ik}$ is the $ik$th element of matrix ${\bf \bar{L}}^{-1}$. Previous work~\cite{FoPiReSa07} has reported that $\bar{l}^{-1}_{ik}$ can be represented by the entries of the pseudoinverse of the Laplacian matrix ${\bf L}^{+}$ as
\begin{equation}\label{EE07}
\bar{l}^{-1}_{ik}=l_{ik}^{+}-l_{iN}^{+}-l_{Nk}^{+}+l_{NN}^{+}\,,
\end{equation}
where the four terms on the right-hand side of equal mark are four entries of ${\bf L}^{+}$.

Equations~(\ref{EE06}) and~(\ref{EE07}) show that in order to determine $T_{iN}$, one can alternatively evaluate the entries of the pseudoinverse matrix ${\bf L}^{+}$. It is known that the elements $l_{ij}^{+}$ of ${\bf L}^{+}$ can be expressed in terms of the eigenvalues and their normalized eigenvectors of ${\bf L}$~\cite{BeGr74, CaMe79}. Let $\lambda_1, \lambda_2, \ldots, \lambda_N$ be the $N$ eigenvalues of $\bf L$, rearranged as $0=\lambda_1<\lambda_2<\cdots<\lambda_N$, and let $\mu_1, \mu_2, \ldots, \mu_N$ be the corresponding mutually orthogonal eigenvectors of unit length, where $\mu_i=(\mu_{i1}, \mu_{i2}, \cdots, \mu_{iN})^{\top}$. According to the spectral graph theory~\cite{Ch97}, ${\bf L}$ has the following spectral form:
\begin{equation}\label{GL01}
{\bf L}=U{\rm diag}\left[\lambda_1, \lambda_2, \cdots, \lambda_N\right] U^{\top},
\end{equation}
where $U=(\mu_1, \mu_2,\cdots, \mu_N)$ is an orthogonal matrix, obeying
\begin{equation}\label{GL02}
UU^{\top}=U^{\top}U=I\,,
\end{equation}
where $I$ is the identity matrix. Thus,
\begin{equation}\label{GL03}
\mu_i^{\top}\mu_j=\delta_{ij},
\end{equation}
where $\delta_{ij}$ is the Kronecker $\delta$ function defined as: $\delta_{ij}=1$ if $i$ equals  $j$, and $\delta_{ij}=0$ otherwise. Education~(\ref{GL01}) also indicates that the entry $l_{ij}$ of matrix $\bf L$ has the following spectral expression:
\begin{equation}\label{GL04}
l_{ij}=\sum_{k=1}^{N}\lambda_{k}\mu_{ki}\mu_{kj},
\end{equation}
which, in the case of $i=j$, reduces to
\begin{equation}\label{GL05}
l_{ii}=s_i=\sum_{k=1}^{N}\lambda_{k}\mu_{ki}^2\,.
\end{equation}

In addition, it is not difficult to find that the entries of eigenvector $\mu_1$ corresponding to $\lambda_1=0$ are all equal to each other, namely $\mu_{1i}=\sqrt N/N$ holds for $1\leq i\leq N$. Hence, for all $1<k\leq N$, the relation
\begin{equation}\label{GL06}
\sum_{i=1}^N \mu_{ki}=0
\end{equation}
holds.

Based on the spectra of $\bf L$, the entries of pseudoinverse ${\bf L}^{+}$ of {\bf L} can be represented as~\cite{BeGr74, CaMe79}
\begin{equation}\label{EE10}
l_{ij}^{+}=\sum_{k=1}^{N}g(\lambda_k)\mu_{ki}\mu_{kj},
\end{equation}
where
\begin{equation}\label{EE11}
g(\lambda_k)=
\begin{cases}
1/\lambda_k,~\lambda_k\neq 0,
\\0,~~~~~~\lambda_k=0.
\end{cases}
\end{equation}

Combining Eq.~(\ref{EE06}), Eq.~(\ref{EE07}), and Eq.~(\ref{EE10}), we can express $T_{iN}$ in terms of the spectra of $\bf L$ as
\begin{eqnarray}\label{EE12}
T_{iN}&=&\sum_{z=1}^{N-1}s_z\sum_{k=2}^{N}\frac{1}{\lambda_k}(\mu_{ki}\mu_{kz}-\mu_{ki}\mu_{kN}-\mu_{kN}\mu_{kz}+\mu_{kN}^2)\nonumber\\
&=&\sum_{z=1}^{N}s_z\sum_{k=2}^{N}\frac{1}{\lambda_k}(\mu_{ki}\mu_{kz}-\mu_{ki}\mu_{kN}-\mu_{kN}\mu_{kz}+\mu_{kN}^2),\nonumber\\
\end{eqnarray}
since for $z=N$ the term $\mu_{ki}\mu_{kz}-\mu_{ki}\mu_{kN}-\mu_{kN}\mu_{kz}+\mu_{kN}^2$ is equal to zero. In the case where $i=N$, Eq.~(\ref{EE12}) indicates $T_{NN}=0$, as suggested by the definition of trapping time. Equation~(\ref{EE12}) is very important since it provides a universal formula for MFPT from one node to another in any weighted network. Note that in a recent paper~\cite{ZhShCh13}, an expression for MFPT was given in terms of the eigenvalues and their corresponding vectors of transition matrix associated with the random walk process, thus Eq.~(\ref{EE12}) is also helpful to unveil the relationship between the spectra of Laplacian matrix and transition matrix of a weighted network.

Inserting Eq.~(\ref{EE12}) into Eq.~(\ref{EEA01}) yields
\begin{eqnarray}\label{EEA04}
T_N&=&\frac{1}{N-1}\sum_{i=1}^N \Bigg[\sum_{z=1}^N s_z\sum_{k=2}^N\frac{1}{\lambda_k}\big(\mu_{ki}\mu_{kz}-\mu_{ki}\mu_{kN}\nonumber\\
&\quad&-\mu_{kN}\mu_{kz}+\mu_{kN}^2\big)\Bigg],
\end{eqnarray}
where we have used the fact of $T_{NN}=0$. Equation~(\ref{EEA04}) can be simplified as
\begin{eqnarray}\label{EEA05}
T_N&=&\frac{1}{N-1}\sum_{k=2}^N\frac{1}{\lambda_k}\Bigg[\sum_{i=1}^N\sum_{z=1}^N s_z\mu_{kN}^2-\sum_{i=1}^N\sum_{z=1}^N s_z\mu_{ki}\mu_{kN}\nonumber\\
&\quad&-\sum_{i=1}^N\sum_{z=1}^N s_z\mu_{kN}\mu_{kz}+\sum_{i=1}^N\sum_{z=1}^N s_z\mu_{ki}\mu_{kz}\bigg].
\end{eqnarray}
The four terms in the square brackets of Eq.~(\ref{EEA05}) can be separately simplified as
\begin{equation}\label{EEA06}
\sum_{i=1}^N\sum_{z=1}^N s_z\mu_{kN}^2=N\times s\times \mu_{kN}^2,
\end{equation}
\begin{equation}\label{EEA07}
\sum_{i=1}^N\sum_{z=1}^N s_z\mu_{ki}\mu_{kN}=s\times\mu_{kN}\sum_{i=1}^N\mu_{ki}=0,
\end{equation}
\begin{equation}\label{EEA08}
\sum_{i=1}^N\sum_{z=1}^N s_z\mu_{kN}\mu_{kz}=N\times\mu_{kN}\sum_{z=1}^N s_z\mu_{kz},
\end{equation}
and
\begin{equation}\label{EEA09}
\sum_{i=1}^N\sum_{z=1}^N s_z\mu_{ki}\mu_{kz}=\sum_{i=1}^N\mu_{ki}\sum_{z=1}^N s_z\mu_{kz}=0,
\end{equation}
where Eq.~(\ref{GL06}) is used. Instituting Eqs.~(\ref{EEA06})-(\ref{EEA09}) into Eq.~(\ref{EEA05}), we obtain an explicit expression for ATT as
\begin{equation}\label{EEA10}
T_N=\frac{N}{N-1}\sum_{k=2}^N\frac{1}{\lambda_k}\left(s\times\mu_{kN}^2-\mu_{kN}\sum_{z=1}^N s_z\mu_{kz}\right),
\end{equation}
which is general valid for the trapping problem with a single trap fixed at any node in an arbitrary weighted network.


\subsection{Lower bound for average trapping time to an arbitrary node}

After obtaining an explicit formula for average trapping time $T_N$ to a given target,  we will continue to deduce a lower bound of it. By Cauchy's inequality, the sum term in Eq.~(\ref{EEA10}) satisfies
\begin{eqnarray}\label{EEB01}
&\quad&\left[\sum_{k=2}^N\frac{1}{\lambda_k}\left(s\times\mu_{kN}^2-\mu_{kN}\sum_{z=1}^N s_z\mu_{kz}\right)\right]\nonumber\\
&\times&\left[\sum_{k=2}^N\lambda_k\left(s\times\mu_{kN}^2-\mu_{kN}\sum_{z=1}^N s_z\mu_{kz}\right)\right]\nonumber\\
&\geq&\left[\sum_{k=2}^N\left(s\times\mu_{kN}^2-\mu_{kN}\sum_{z=1}^N s_z\mu_{kz}\right)\right]^2.
\end{eqnarray}
In Eq.~(\ref{EEB01}), the equality holds if and only if for those $k$ with $s\times\mu_{kN}^2-\mu_{kN}\sum_z s_z\mu_{kz}\neq0$, the corresponding eigenvalues $\lambda_k$ are equal to each other. Considering Eqs.~(\ref{GL04}),~(\ref{GL05}) and $\lambda_1=0$, one obtains
\begin{eqnarray}\label{EEB02}
&\quad&\sum_{k=2}^{N}\lambda_k\left( s\times\mu_{kN}^2-\mu_{kN} \sum_{z=1}^N s_z\mu_{kz}\right)\nonumber\\
&=&s\sum_{k=1}^N \lambda_k\mu_{kN}^2-\sum_{z=1}^N s_z \sum_{k=1}^N \lambda_k\mu_{kN}\mu_{kz}\nonumber\\
&=&s\times s_N-\sum_{z=1}^N s_z l_{zN}=s_N(s-s_N)+\sum_{z\neq N}s_z w_{Nz}.\nonumber\\
\end{eqnarray}
Moreover,
\begin{eqnarray}\label{EEB03}
&\quad&\sum_{k=2}^N\left(s\times\mu_{kN}^2-\mu_{kN}\sum_{z=1}^N s_z\mu_{kz}\right)\nonumber\\
&=&\left(s\sum_{k=1}^N \mu_{kN}^2-s\mu_{1N}^2\right)\nonumber\\
&\quad&-\left(\sum_{z=1}^N s_z\sum_{k=1}^N\mu_{kN}\mu_{kz}-\mu_{1j}\sum_{z=1}^N s_z\mu_{1z}\right)\nonumber\\
&=&s-\frac{s}{N}-s_N+\frac{s}{N}=s-s_N,
\end{eqnarray}
where Eq.~(\ref{GL03}) and $\mu_{1i}=\sqrt{N}/N$ are used.
Combining the above obtained results yields
\begin{equation}\label{EEB04}
T_N\geq\frac{N}{N-1}\frac{(s-s_N)^2}{s_N(s-s_N)+\sum_{z\neq N}s_z w_{Nz}}.
\end{equation}
In this way, we have derived a lower bound for ATT, which is expressed in terms of the fundamental parameters of a weighted network.

The lower bound provided in Eq.~(\ref{EEB04}) is sharp since it can be reached in some graphs. For instance, for trapping on those networks where all nodes are connected to the trap by edges with equal weight, this lower bound can be achieved, the reason of which can be explained as follows. By construction, in these networks the weight of any edge connecting a non-trap node $i (1\leq i \leq N-1)$ and the trap node $N$ is $w_{iN}=w_{N i}=s_N/(N-1)$. Then, the row vector corresponding to the $N$th row of the Laplacian matrix is
\begin{equation}
\left(-\frac{s_N}{N-1}, -\frac{s_N}{N-1}, \cdots, -\frac{s_N}{N-1}, s_N\right),
\end{equation}
which multiplying $\mu_i$ ($ 1\leq i \leq N-1$) leads to
\begin{equation}\label{EEB05}
-\frac{s_N}{N-1}\sum_{k=1}^{N-1}\mu_{ik}+s_N\mu_{iN}=\lambda_i\mu_{iN}.
\end{equation}
Recalling Eq.~(\ref{GL06}), we have $\sum_{k=1}^{N-1}\mu_{ik}=-\mu_{iN}$. Thus, Eq.~(\ref{EEB05}) can be recast as
\begin{equation}\label{EEB06}
\frac{N}{N-1}s_N\mu_{iN}=\lambda_i\mu_{iN}.
\end{equation}
Notice that Eq.~(\ref{EEB06}) holds if and only if one of two conditions are satisfied: (i) $\mu_{iN}=0$;  (ii) $\mu_{iN} \neq 0$ but $\lambda_i=s_N N/(N-1)$ ($ 1\leq i \leq N-1$).
These satisfy the condition under which the equality of Eq.~(\ref{EEB01}) holds. Therefore, for this group of networks, when the trap is located at a node linked to all other nodes by edges with equal weight, the ATT is exactly the lower bound given by Eq.~(\ref{EEB04}),
\begin{eqnarray}\label{EEB07}
T_N&=&\frac{N}{N-1}\frac{(s-s_N)^2}{s_N(s-s_N)+\frac{s_N}{N-1}\sum_{z\neq N}s_z}\nonumber\\
&=&\frac{N}{N-1}\frac{(s-s_N)^2}{s_N(s-s_N)+\frac{s_N}{N-1}(s-s_N)}\nonumber\\
&=&\frac{s-s_N}{s_N},
\end{eqnarray}
which is dependent only on the total strength and the strength of the trap node. It is obvious that the lower bound in Eq.~(\ref{EEB04}) can be reached in both the binary complete graph and the star graph, when the trap is placed at a most connected node.



In general large weighted networks, especially real-life networks~\cite{BoLaMoChHw06}, the strength of an individual node is much smaller compared with the sum of strength of all nodes, it is the same with the weight of an edge. Thus, the leading term of Eq.~(\ref{EEB04}) is
\begin{equation}\label{EEB09}
T_N \simeq \frac{(s-s_N)^2}{s_N(s-s_N)}\sim \frac{s}{s_N},
\end{equation}
which is proportional to the total strength and the inverse strength of the target node.

The obtained lower bound for ATT given in Eq.~(\ref{EEB04}) provides important information about trapping in a weighted graph since it establishes a range of ATT in terms of simple parameters of the graph, including the number of nodes, total strength of all nodes, strength of those nodes adjacent to the trap, and weight of edges incident to the trap node. Furthermore, the lower bound can only be attained in serval particular graphs, but it's leading term, given in Eq.~(\ref{EEB09}), can be obtained in uncorrelated weighted graphs, a special class of which can mimic the edge weights of a wide range of real-world networks~\cite{BaBaPaVe04,MaAlBa05}, as will be addressed in Sec.~\ref{uncorre}.

It should be stressed that in a binary network, the ATT to a target and its bound depend on the structure of the whole network~\cite{TeBeVo09}. However, for weighted networks, the case differs substantially. In addition to network structure, weight distribution also plays a significant role in the ATT. For example, for trapping on the binary complete graph with a trap positioned at an arbitrary node, the minimal ATT can be attained~\cite{Bobe05}. But for trapping on weighted complete graphs, the weight distribution strongly influences the ATT: Different weight distribution could make the ATT close to or deviate from the lower bound.

\subsection{Average trapping time for trapping with the trap uniformly distributed
\label{TrapUnif}}

In the above, we have considered the trapping problem on weighted networks with a single trap placed at any given position. Here, we address another trapping problem on a weighted network
with a trap uniformly distributed over the whole network. In this case, the ATT, denoted by $\langle T \rangle$, is defined as the average of MFPTs over all pairs of nodes in the networks.
Thus, the quantity $\langle T \rangle$ involves a double average. The former is over all the source nodes to a given target node, the latter is the average of the first one.
That is,
\begin{equation}\label{EEC01}
\langle T\rangle=\frac{1}{N(N-1)}\sum_{j=1} ^N \sum_{\substack{i=1 \\ i \neq j}} ^N T_{ij}=\frac{1}{N}\sum_{j=1} ^N T_j.
\end{equation}
Plugging Eq.~(\ref{EEA10}) into Eq.~(\ref{EEC01}), we have
\begin{equation}\label{EEC02}
\langle T\rangle=\frac{1}{N-1}\sum_{k=2}^{N}\frac{1}{\lambda_k}\left(s \sum_{j=1}^N \mu_{kj}^2-\sum_{j=1}^N \mu_{kj}\sum_{z=1}^N s_z\mu_{kj}\mu_{kz}\right).
\end{equation}
Considering $\sum_{j=1}^N \mu_{kj}^2=1$ and $\sum_{j=1}^N \mu_{kj}=0$, the expression for $\langle T\rangle$ is reduced to
\begin{equation}\label{EEC03}
\langle T\rangle=\frac{s}{N-1}\sum_{k=2}^N\frac{1}{\lambda_k}\,,
\end{equation}
which depends on the number of nodes, sum of strengths of all nodes, and all nonzero eigenvalues of Laplacian matrix $\bf L$.


Hitherto, we have studied two cases of trapping problems: trapping with a fixed trap and trapping with a trap uniformly distributed among all nodes, and obtained the exact formulas for quantities $T_{N}$ and $\langle T\rangle$, both of which are general and applicable to all weighted connected networks.

In the sequel, we will utilize our theoretical results to analyze the trapping problems on two particulary networks. We first consider random walks on weighted uncorrelated networks with a single trap, focusing on a network family whose weights are controlled by a tunable parameter $\theta$. After that, we will study L\'evy random walks in connected binary networks, which can be described by random walks on weighted networks derived from their corresponding original binary networks, concentrating on the case that the trap is uniformly distributed. For both trapping problems, we will show that the weights have substantial influence on the trapping efficiency measured by ATT.

\section{Trapping in weighted uncorrelated networks with an immobile trap \label{uncorre}}

In this section, we apply the above-obtained results to uncorrelated networks having a single trap located at an arbitrary node. We concentrate on weighted networks with a specific form of the weights of edges. In these networks, the weight $w_{ij}$ of an edge connecting nodes $i$ and $j$ is $w_{ij}=(d_i d_j)^{\theta}$, where $\theta$ is a controllable parameter and $d_i$ and $d_j$ are the degrees of $i$ and $j$, respectively.
Such a weight-assigning form of weighted networks is typical of diverse real systems, such as the airport networks~\cite{BaBaPaVe04,MaAlBa05}, scientific collaboration networks~\cite{BaBaPaVe04}, and metabolic networks~\cite{MaAlBa05}.


\begin{figure*}
\centering
$\begin{array}{cc}
{
\includegraphics[width=0.34\linewidth,trim=60 40 60 40]{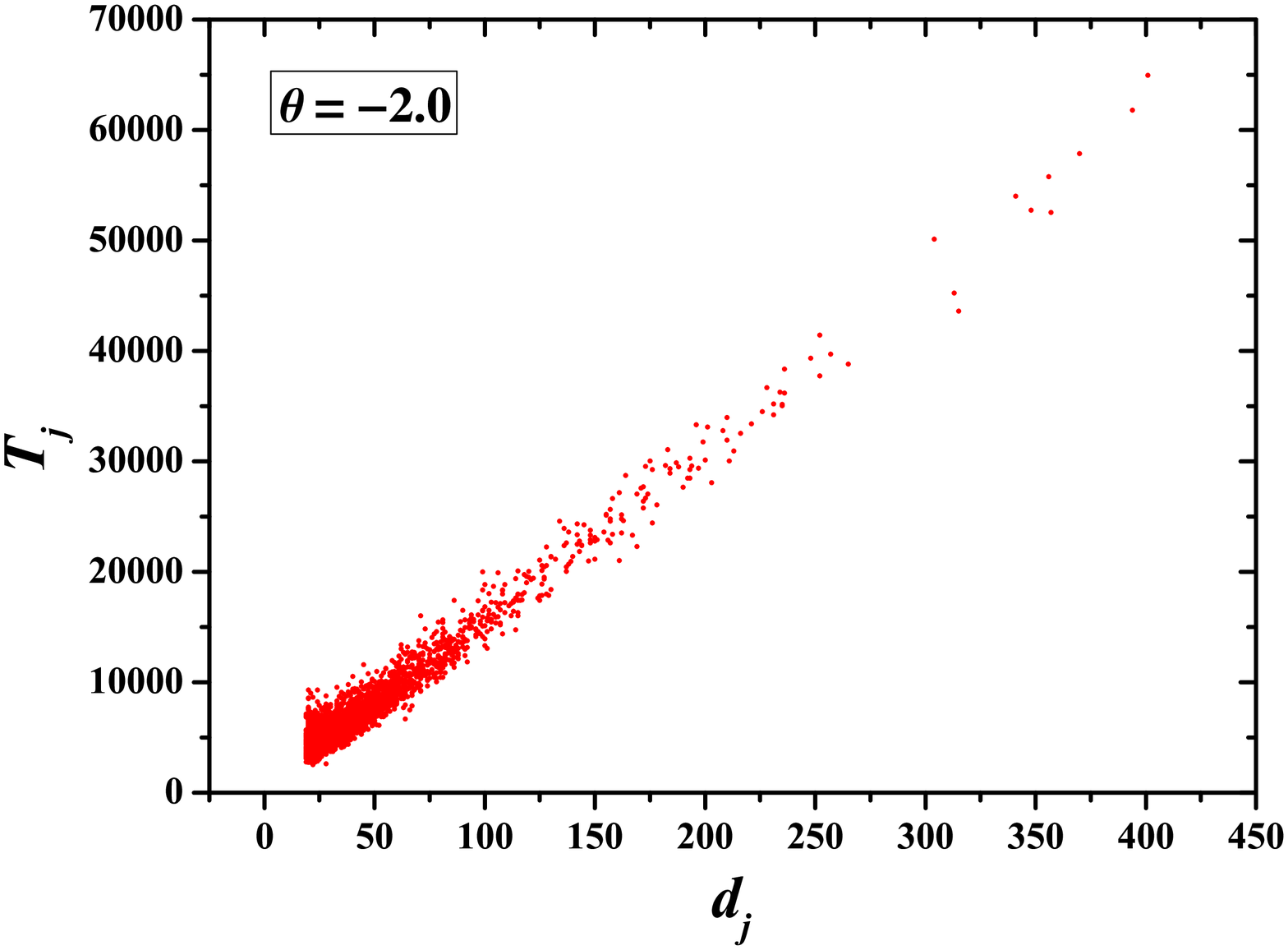}
}
{
\includegraphics[width=0.34\linewidth,trim=60 40 60 40]{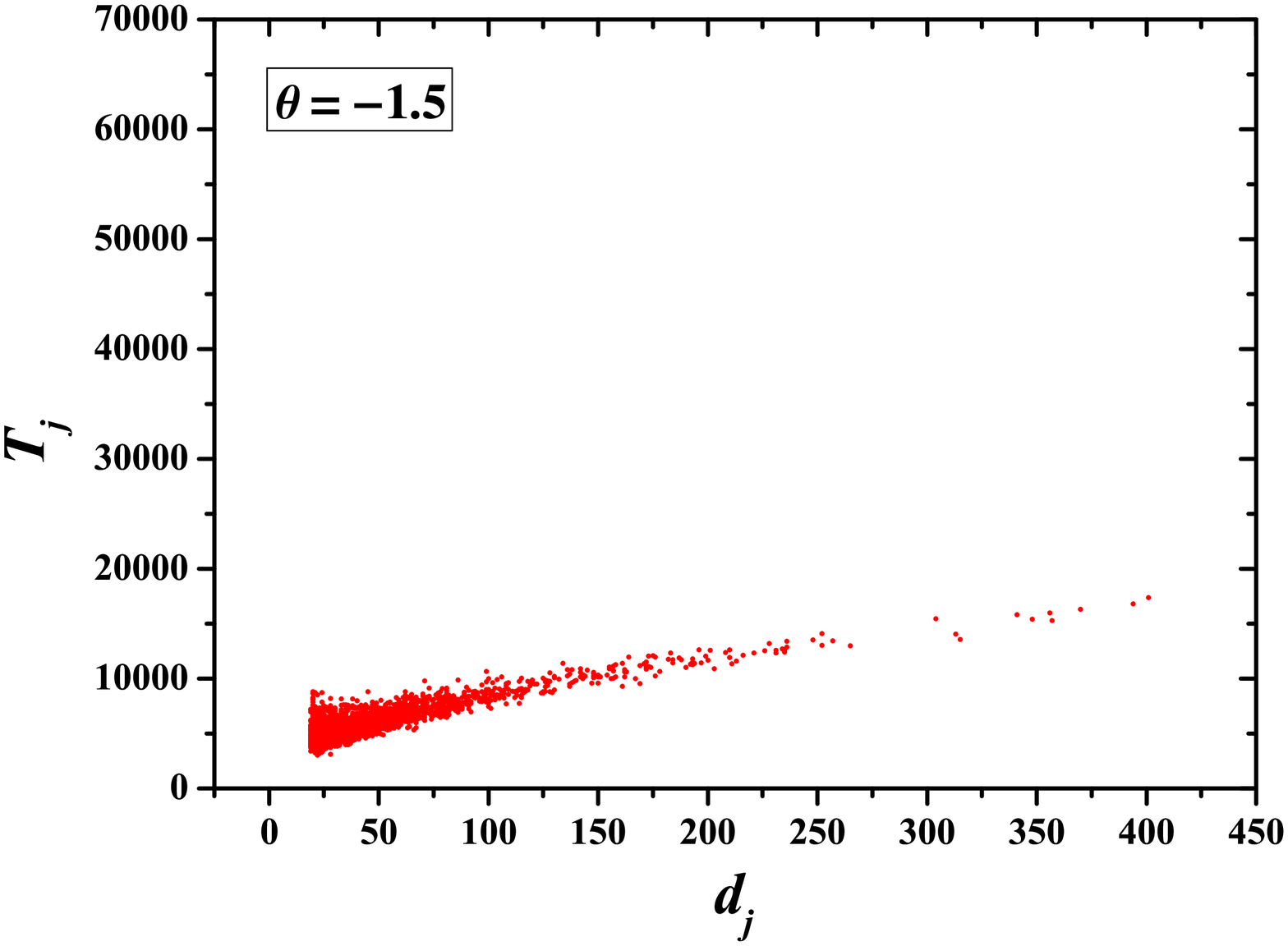}
}
{
\includegraphics[width=0.34\linewidth,trim=60 40 60 40]{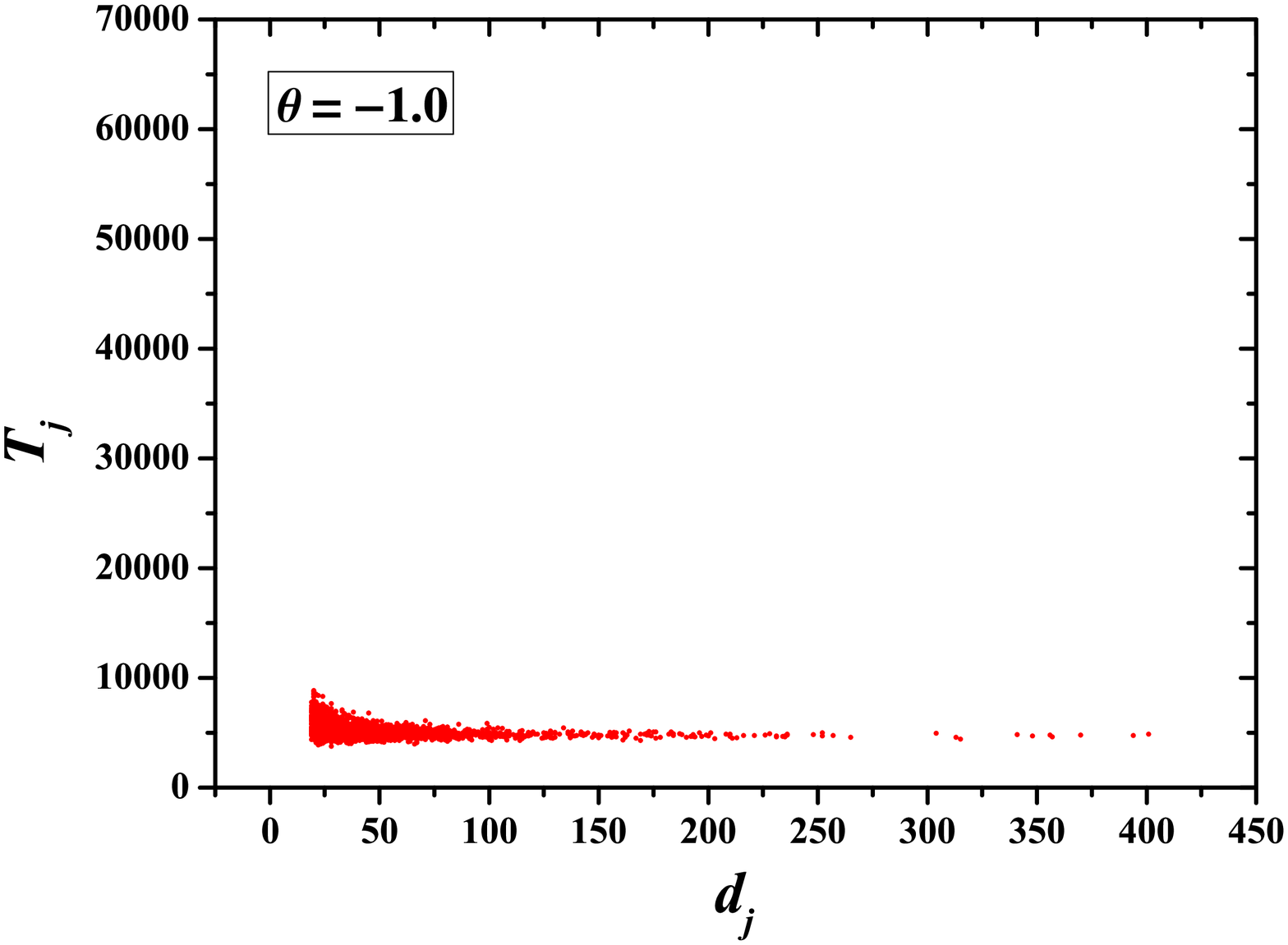}
}\\
\\
{
\includegraphics[width=0.34\linewidth,trim=60 40 60 40]{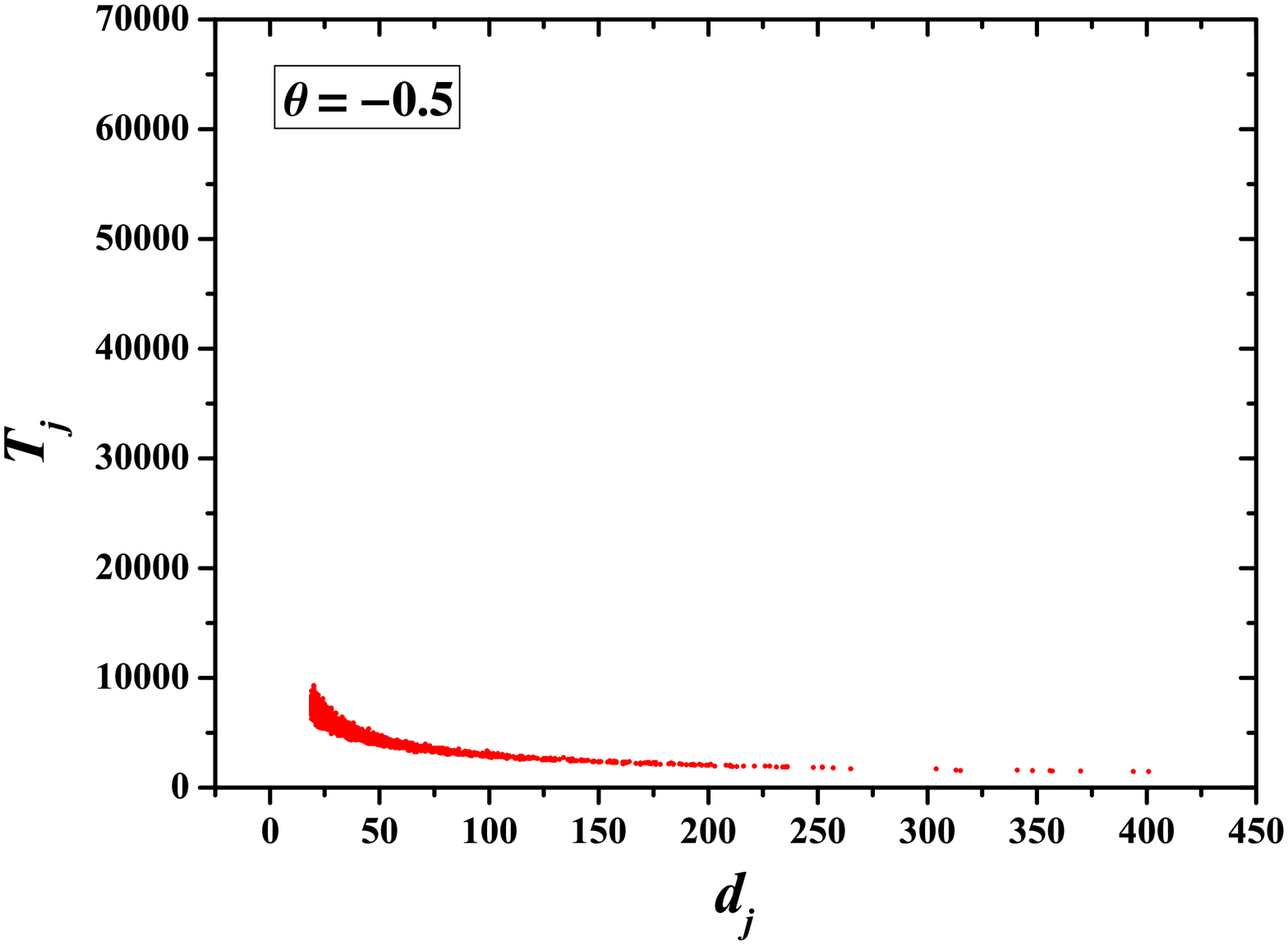}
}
{
\includegraphics[width=0.34\linewidth,trim=60 40 60 40]{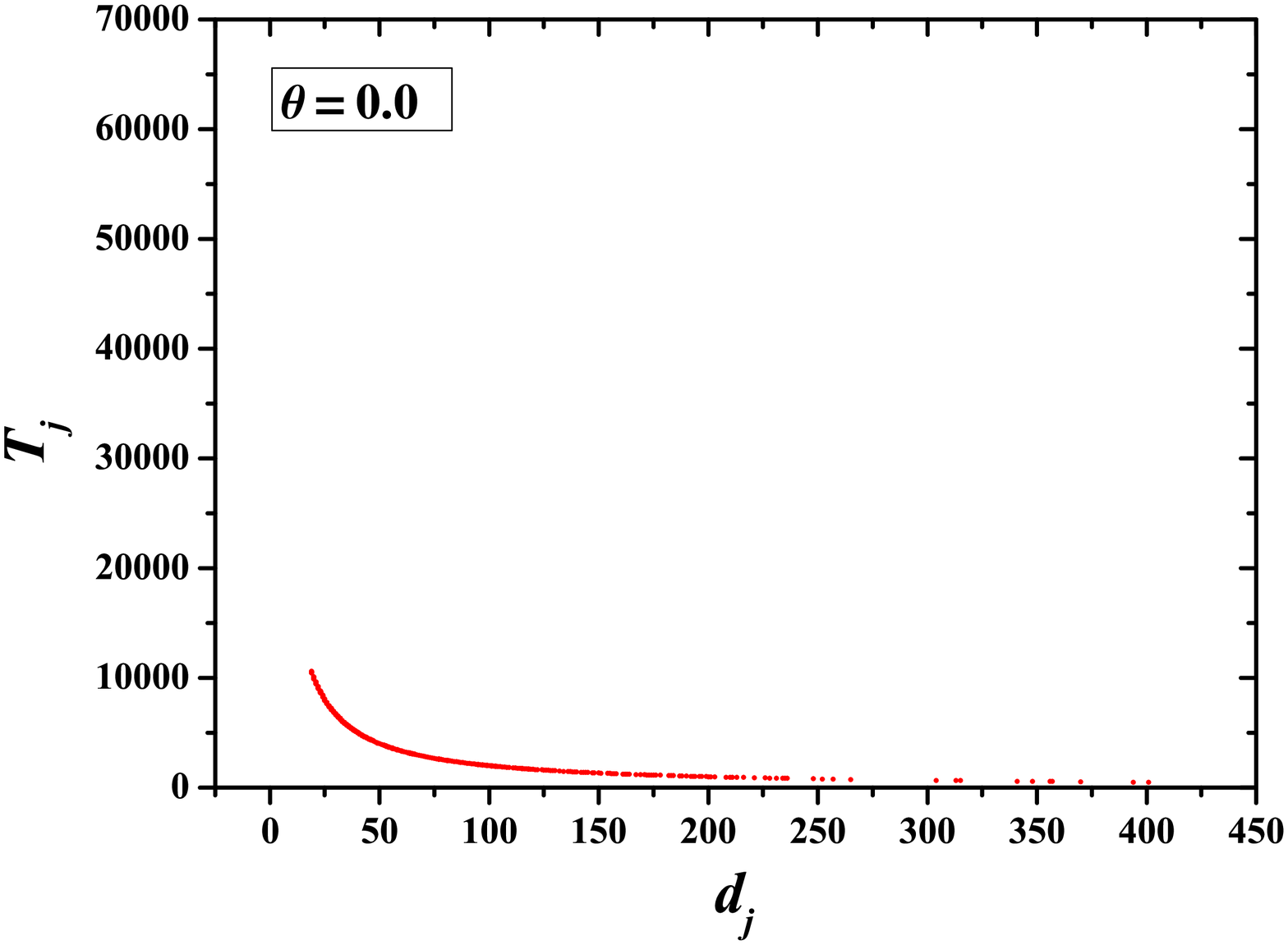}
}
{
\includegraphics[width=0.34\linewidth,trim=60 40 60 40]{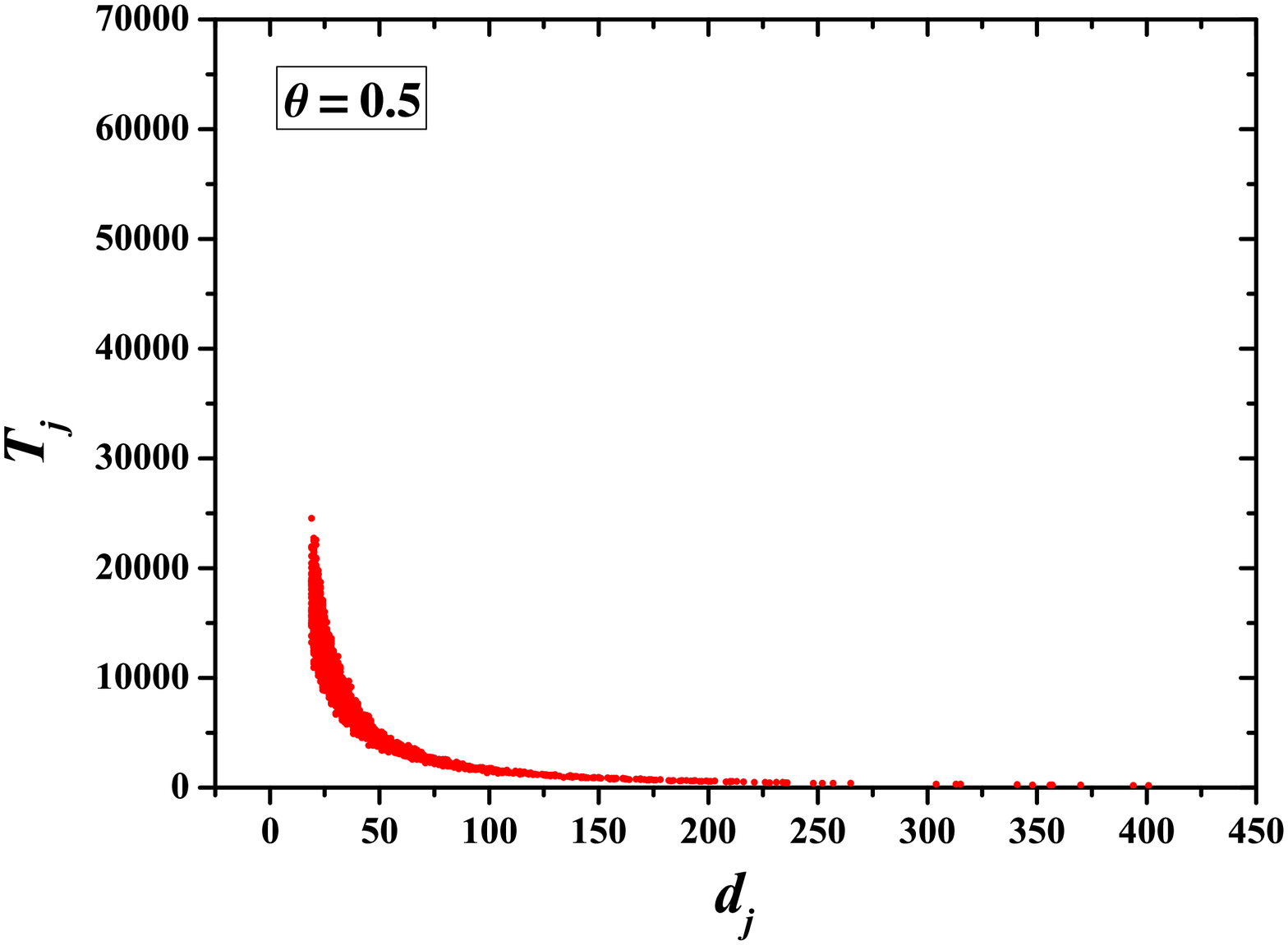}
}
\end{array}$
\caption[kurzform]{(Color online) The changes of average trapping time with weight parameter $\theta$ and the degree of the target nodes in a BA network with size $N=5000$ and $\langle d\rangle=40$.}\label{ATT6}
\end{figure*}

Since the weight exponent $\theta$ characterizes the weights of different networks~\cite{PoStZl12}, where the strength $s_i$ of node $i$ depends on its degree $d_i$ satisfying $s_i \sim d_i^\beta$ with $\beta=\theta+1$, our main goal is to unveil the relationship between the weight exponent $\theta$ and the ATT to different nodes, and to better understand how the weights affect the behavior of the trapping process. To this end, we utilize Eq.~(\ref{EEA10}) to perform numerical simulations on the Barab\'asi-Albert (BA) network~\cite{BaAl99}, which is scale-free having a power-law degree distribution $P(d)\sim d^{-\gamma}$ with the exponent $\gamma=3$. In Fig.~\ref{ATT6}, we report the exact numerical results for ATTs to nodes with different degrees. From Fig.~\ref{ATT6}, it is found that the weight parameter $\theta$ has a strong and different influence on the ATT to target nodes of disparate degrees. For $\theta <-1.0$, the ATT grows with the increase of node degree; for $\theta=-1.0$, the ATT for all nodes is approximately equal to each other; while for $\theta >-1.0$, the ATT is a decreasing function of trap's degree: The larger the degree of trap node, the smaller the ATT. These rich phenomena are in marked contrast to that observed in the binary BA networks corresponding to the case of $\theta=0$~\cite{KiCaHaAr08}.

The observed rich behaviors can be understood based on the following theoretical analysis that is applicable to all uncorrelated networks, including those being studied. For trapping process in a network, binary or weighted, a basic quantity characterizing this dynamical process is survival probability, $\rho(t)$, defined as the probability that a particle survives after $t$ jumping steps~\cite{BaKl98JPC,KiCaHaAr08}. The ATT $T_j$ then can be represented in terms of $\rho(t)$ as
\begin{equation}\label{TT01}
T_j=\sum_{t=0}^{\infty}\rho(t).
\end{equation}
For random walks in a completely uncorrelated weighted network, at every time step the expected probability that a walker jumps from its current location to the target node $j$ is proportional to the relative strength of the target node. Thus, the discrete time master equation describing survival probability is
\begin{equation}\label{TT02}
\rho(t+1)=\rho(t)\left(1-\frac{s_j}{s}\right).
\end{equation}
Considering $\rho(0)=1$,  Eq.~(\ref{TT02}) is easy to solved to yield
\begin{equation}\label{TT03}
\rho(t)=\left(\frac{s-s_j}{s}\right)^t.
\end{equation}
Plugging Eq.~(\ref{TT03}) into Eq.~(\ref{TT01}), the ATT to the trap node $j$ can be estimated as
\begin{equation}\label{AT02}
T_j=\sum_{t=0}^\infty\left(\frac{s-s_j}{s}\right)^t=\frac{s}{s_j}.
\end{equation}

Equation~(\ref{AT02}) is consistent with Eq.~(\ref{EEB09}), indicating that for trapping on an uncorrelated weighted network, the possible minimal scaling for ATT can be achieved when the trap is placed at any node on the network. Thus, the trapping process in uncorrelated networks is very efficient.

For the special class of weighted networks with weight $w_{ij}=(d_i d_j)^{\theta}$ for the edge connecting nodes $i$ and $j$, many related quantities can be determined explicitly. For a node $i$, its strength $s_i$ can be evaluated as
\begin{equation}\label{AT03}
s_i=\sum_{j=1}^{N}(d_i d_j)^{\theta}=(d_i)^{\theta}\sum_{d'=d_{\rm min}}^{d_{\rm max}}d_i P(d'|d_i)(d')^{\theta},
\end{equation}
where $P(d'|d_i)$ is the conditional probability~\cite{PaVaVe01} that a node of degree $d_i$ is directly to a node with degree $d'$; $d_{\rm min}$ and $d_{\rm max}$ represent the minimum and maximum node degrees, respectively.

In uncorrelated networks, the degrees of the two nodes connecting any edge are completely independent. The conditional probability $P(d'|d_i)$ then can be simply
estimated as $P(d'|d_i)=d'P(d')/\langle d \rangle$, where $P(d)$ is the degree distribution and $\langle d\rangle$ is the average node degree. Therefore,
\begin{equation}\label{AT04}
s_i=(d_i)^{\theta+1}\sum_{d'=d_{\rm min}}^{d_{\rm max}}\frac{(d')^{\theta+1}P(d')}{\langle d\rangle}=\frac{(d_i)^{\theta+1}\langle d^{\theta+1}\rangle}{\langle d\rangle},
\end{equation}
where $\langle d^{\theta+1}\rangle$ is the ($\theta+1$)th-order moment of the degree distribution. Using Eq.~(\ref{AT04}), we can further evaluate the sum of strengths of all nodes as
\begin{equation}\label{AT05}
s=\sum_{i}NP(d_i)s_i=\frac{N\langle d^{\theta+1}\rangle^2}{\langle d\rangle}.
\end{equation}

Substituting the results of Eqs.~(\ref{AT04}) and~(\ref{AT05}) into Eq.~(\ref{AT02}) yields
\begin{equation}\label{AT06}
T_j=\frac{N\langle d^{\theta+1}\rangle}{(d_j)^{\theta+1}}\,,
\end{equation}
which provides an approximation of ATT to any target node.
This formula shows that for $\theta<-1$ ($\theta>-1$), the ATT is an increasing (a decreasing) function of degree of the trap node. While for $\theta=-1$, the ATT becomes a constant, irrespective of the trap's degree. Therefore, the analytical prediction in Eq.~(\ref{AT06}) provides not only an interpretation for the phenomena in Fig.~\ref{ATT6} but also an effective
method to control the efficiency of trapping in uncorrelated weighted networks, the latter of which has become an outstanding issue in the area of complex
systems~\cite{LiSlBa11}.


\section{Trapping of L\'evy random walks with trap distributed uniformly}



The L\'evy random walks~\cite{RiMa12} are inspired by L\'evy flights~\cite{MeKl04} where the random displacements $l$ obey asymptotically a power-law probability distribution $P(l)\sim l^{-\alpha}$. L\'evy flights exhibit a better performance in various problems, such as human mobility and behavior~\cite{BrHuGe06,BrLiGl07,RhShHoLeKiCh11,Sc11,RaBaAm12,RaBa12} and foraging by animals~\cite{ViLuRaSt11, Si08, JaWeHeNoKo11}. It was reported recently that long-range navigation based on the L\'evy random walk strategy is more efficient than that using the standard random walk strategy~\cite{RiMa12}. Here we are mainly concerned with the behavior of the process of L\'evy random walks in a binary network with a trap distributed uniformly in the whole network.

During the process of L\'evy random walks~\cite{RiMa12}, at each time, a walker located at a current node $i$ can jump to any other node, say $j$, with a probability proportional to $r_{ij}^{-\alpha}$, where $r_{ij}$ is the minimum distance between nodes $i$ and $j$ and $\alpha$ is a nonnegative controllable parameter. Evidently, the L\'evy random walks in a binary network are biased, which are identical to standard random walks in a corresponding weighted complete graph~\cite{LaSiDeEvBaLa11}, where the weight of edge connecting two nodes $i$ and $j$ is $w_{ij}=r_{ij}^{-\alpha}$. The equivalence between these two random walks allows to use the obtained result in Sec.~\ref{TrapUnif} to analyze $\langle T \rangle$ of the L\'evy random walks.

For weighted networks with edge weight $w_{ij}=r_{ij}^{-\alpha}$, the strength of node $i$ is
\begin{equation}\label{GA01}
s_i=\sum_{\substack{j=1 \\ j \neq i}} ^N r_{ij}^{-\alpha},
\end{equation}
and the total strength is
\begin{equation}\label{GA02}
s=\sum_{i=1}^N s_i = \sum_i^N \sum_{\substack{j=1 \\ j \neq i}}^N r_{ij}^{-\alpha}.
\end{equation}
For random walks on such weighted networks, the transition probability from node $i$ to node $j$ is $r_{ij}^{-\alpha}/s_i$. The random walks have two limited cases: $\alpha\to +\infty$ and $\alpha=0$. The former corresponds to the unbiased walks in the original binary networks, while the latter is exactly the unbiased walks in unweighted complete graphs with the same size as the original binary networks.

\begin{figure}
\begin{center}
\includegraphics[width=1.1 \linewidth,trim=50 70 0 50]{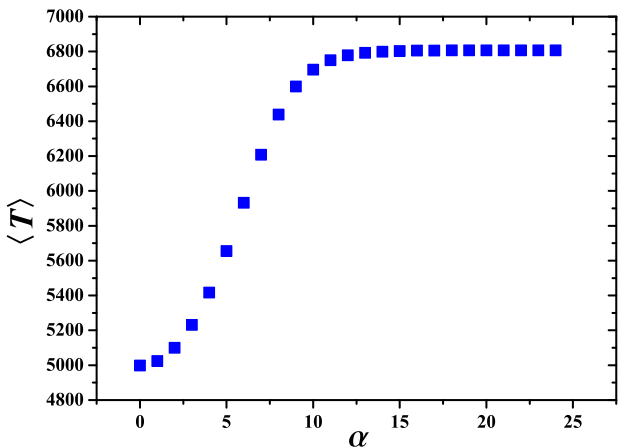}
\end{center}
\caption[kurzform]{(Color online) The effect of  parameter $\alpha$ on $\langle T \rangle$ for L\'evy random walks on a BA network with $N=5000$ and $\langle d \rangle=40$. }\label{GATT}
\end{figure}

After reducing the problem of L\'evy random walks in binary graphs to that of standard random walks in corresponding weighted networks, we are in a position to study the average of MFPTs over all pairs of nodes by using the general result derived in Sec.~\ref{TrapUnif}. According to Eq.~(\ref{EEC03}), we compute $\langle T\rangle$ for L\'evy random walks in a BA scale-free network with size $N=5000$. The exact numerical results are reported in Fig.~\ref{GATT}, which shows that $\langle T \rangle$ increases monotonously with parameter $\alpha$. When $\alpha=0$, $\langle T \rangle$ reaches its minimal value $N-1$; when $\alpha$ grows from zero to $\infty$,
$\langle T \rangle$ increases from $N-1$ and tends to a constant for large $\alpha$. The minimal value occurring at $\alpha=0$ implies that uniform distribution of edge weights is the optimal distribution of weights in the sense that it minimizes $\langle T \rangle$.

The result that $\alpha=0$ is the optimal parameter minimizing $\langle T \rangle$ can be proved by using the method of Lagrange multipliers.
Let's examine the following function:
\begin{equation}\label{Optima01}
f(\lambda_2,\cdots,\lambda_N) = \sum_{k=2}^{N}\frac{s}{\lambda_{k}}\,,
\end{equation}
subject to the conditions of $\lambda_k>0$ ($2\leq k \leq N$) and $\lambda_2+\lambda_3+\cdots +\lambda_N=s$. Applying the method of Lagrange multipliers to Eq.~(\ref{Optima01}), the minimum of the function $f(\lambda_2,\cdots,\lambda_N)$ is attained at $\lambda_2 = \lambda_3 = \cdots = \lambda_N = s/(N-1)$. Thus, the minimum $\langle T \rangle$ among all weighted networks of size $N$ is
\begin{equation}\label{Optima02}
\langle T \rangle_{\rm min} =\frac{s}{N-1}\sum_{k=2}^{N}\frac{1}{\lambda_{k}}=N-1\,.
\end{equation}

Using the technique in Ref.~\cite{Zh08}, it is easy to prove that, the above set of identical eigenvalues $\lambda_2 = \lambda_3 = \ldots = \lambda_N =s/(N-1)$ corresponds to those of the Laplacian matrix of a family of complete graphs with $N$ nodes, in which every node is connected to all others by links with the same weight $s/[N(N-1)]$. This graph family includes the binary complete graph associated with $\alpha=0$ as a particular case, where all the $N(N-1)/2$ edges have a unit weight, while the weighted complete graphs corresponding to other values of $\alpha$ are not subsumed. Thus, for L\'evy random walks in a connected undirected network, the case of $\alpha=0$ is optimal in the sense that it has the absolute minimum value for $\langle T \rangle$.

\section{conclusions}

In this paper, we have developed a universal framework for random walks, especially trapping problem, in a weighted network. We have derived an analytical expression for MFPT from one node to another in terms of the eigenvalues and eigenvectors of Laplacian matrix for the network, and obtained an explicit solution to ATT for trapping in the network with a deep trap placed at any node. By employing Cauchy's inequality, we have provided a lower bound for ATT that is expressed only by the local information of the trap node. The lower bound is sharp and can be attained in those networks, where the trap is connected to all other nodes by links with equal weight. We have further deduced an exact formula for the average of MFPTs in terms of only the eigenvalues, where the average is taken over all node pairs in the network. All obtained results indicate that the network weights have a significant influence on the trapping process.

To show the generality of our framework and methods, we have studied two trapping problems in weighted networks. The first trapping problem is random walks in a family of uncorrelated weighted networks with a single trap located at any node, where the weights are dominated by a parameter; the other trapping issue is L\'evy random walks (characterized by an exponent) in a connected binary network with the trap uniformly distributed in the whole network, which can be considered as trapping in a corresponding weighted network. For trapping in uncorrelated weighted networks, we have shown the ATT to any trap can display various behaviors, determined by the weight parameter. Although the ATT cannot attain the possible maximal lower bound, its leading scaling is the same as that of the lower bound. For L\'evy random walks with a uniform-distribution trap, we have demonstrated that the trapping process is the most efficient when the characteristic exponent is zero.

Since various types of biased random walks (e.g., maximal-entropy random walks~\cite{GaLa08,BuDuLuWa09,SiGoLaNiLa11}) in binary networks can be represented as random walks in weighted networks~\cite{LaSiDeEvBaLa11}, our theoretical framework proves to be a powerful tool to attack various random walks in binary and weighted networks, improving our understanding of random walks and trapping process. Moreover, this work provides a novel spectral viewpoint to study random walks and trapping in binary and weighted networks in terms of the graph Laplacian.

\begin{acknowledgments}
This work was supported by the National Natural Science Foundation of China under Grants No. 61074119 and No. 11275049.
\end{acknowledgments}



\end{document}